\documentclass[12pt]{article}

\usepackage{paper2e}
\usepackage{mydefs2e}
\usepackage{xspace}
\usepackage{epsfig}

\renewcommand{\d}{\partial}

\begin{document}

\begin{titlepage}
\preprint{UMD-PP-06-007}

\title{Mixed Gauge and Anomaly Mediation\\\medskip
From New Physics at 10 TeV}

\author{Ken Hsieh,\ \ Markus A. Luty}

\address{Physics Department, University of Maryland\\
College Park, Maryland 20742}

\begin{abstract}
In the context of anomaly-mediated supersymmetry breaking,
it is natural for vectorlike fields and singlets
to have supersymmetry breaking masses of order 10~TeV,
and therefore act as messengers of supersymmetry breaking.
We show that this can give rise to phenomenologically
viable spectra compatible with perturbative gauge coupling
unification.
The minimal model interpolates continuously between pure
anomaly mediation and gauge mediation with a messenger
scale of order $10\TeV$.
It is also possible to have non-minimal models with more
degenerate specta, with some squarks
lighter than sleptons.
These models reduce to the MSSM at
low energies and incorporate a natural solution of the $\mu$ problem.
The minimal model has four continuous parameters and one discrete
parameter (the number of messengers).
The LEP Higgs mass bound can be satisfied in the minimal model
by tuning parameters at the GUT scale to one part in $50$.
\end{abstract}

\end{titlepage}

\section{Introduction}
\label{sec:intro}
\newcommand{\Pt}{\tilde{P}}
Anomaly-mediated supersymmetry breaking (AMSB) \cite{AMSB,AMSB2}
is an attractive mechanism for breaking supersymmetry (SUSY) without
flavor problems.
In this mechanism, SUSY is broken by the VEV of a supergravity
auxiliary field $\avg{F_\phi}$, whose couplings to matter are
governed by scale covariance, and are hence naturally
flavor-blind.
It defines a preferred renormalization group (RG) trajectory
for all SUSY breaking couplings in terms of a single
SUSY breaking scale $\avg{F_\phi} \sim 10\TeV$.
Unfortunately, the slepton mass parameters are negative in the
minimal supersymmetric standard model (MSSM).
In this paper, we propose a solution to this problem
based on an idea due to Nelson and Weiner
\cite{NW}, which built on early work by Pomarol and Rattazzi \cite{PR}.
Nelson and Weiner considered a theory with extra vectorlike fields
$P$ and $\Pt$ and added a coupling of the form%
\footnote{Couplings of this form with $P$ and $\tilde{P}$ replaced by the
MSSM Higgs fields contribute to the Giudice-Masiero mechanism for generating
the MSSM $\mu$ term \cite{GM}.}
\beq
\De \scr{L} &= \myint d^4\th\, \frac{\phi}{\phi^\dagger} c P \Pt
+ \hc
\eeq
This gives rise to a Dirac fermion mass $c \avg{F_\phi}$ and scalar
mass terms
\beq
V = | c \avg{F_\phi} |^2 (|P|^2 + |\Pt|^2)
+ \left( c |\avg{F_\phi}|^2 P \Pt + \hc \right).
\eeq
The scalar mass-squared terms are positive for $|c| > 1$.
Assuming $|c| \sim 1$, this is a supersymmetry
breaking threshold at the scale $\avg{F_\phi}$,
which gives SUSY breaking threshold corrections of order
$g^2 \avg{F_\phi} / 16\pi^2$ to SUSY breaking masses, taking them
off the AMSB RG trajectory.
As shown in \Ref{NW}, the leading threshold corrections to the scalar masses
vanish,  and the slepton mass-squared terms are therefore still negative
at the scale $\avg{F_\phi}$.
One can get positive slepton masses at the weak scale
only by having a large number
of messengers (5 or more ${\bf 5} \oplus \bar{\bf 5}$'s),
which generates large gaugino masses at the messenger scale $\sim 10\TeV$,
which in turn generates positive slepton masses from running
between the messenger scale and the weak scale.
However, the resulting theories generally have charged slepton
LSP, and the large number of messengers destroys perturbative
unification.

In this paper we consider a very simple extension of this model
that has a more attractive phenomenology.
The model consists of the MSSM plus a singlet $S$ in addition to
the vectorlike fields $P$, $\Pt$.
We include the most general interactions with dimensionless
coefficients.
The additional terms in the Lagrangian are therefore
\beq[Ladd]
\bal
\De\scr{L} &= \myint d^4\th \frac{\phi^\dagger}{\phi} \left(
\sfrac 12 c_S S^2 + c_P P \Pt \right) + \hc
\\
&\qquad + \myint d^2\th \left[
\frac{\la_S}{3!} S^3 + \la_P S P \Pt \right] + \hc
\eal\eeq
A superpotential coupling of the form $S H_u H_d$ is assumed to
be absent.%
\footnote{For example, it may be forbidden by a discrete $R$ symmetry
$S(\th) \mapsto -S(i\th)$,
$P(\th) \mapsto +P(i\th)$,
$\tilde{P}(\th) \mapsto +\tilde{P}(i\th)$,
$H_u(i\th) \mapsto +H_u(i\th)$,
$H_d(i\th) \mapsto -H_d(i\th)$,
$u^{\rm c}(\th) \mapsto -u^{\rm c}(i\th)$,
with all other fields even.}
For $|c_S| < 1$ the potential for $S$ has a local maximum at
$S = 0$, so $\avg{S} \ne 0$.
This gives rise to a more general threshold with none of the problems
of the minimal model.

\section{The Threshold}
\label{sec:model}
In this section, we compute the SUSY breaking from the threshold.
The scalar potential that arises from \Eq{Ladd} is
\beq[theVS]
\bal
V &= \left| c_S \avg{F_\phi^\dagger} S + \sfrac 12 \la_S S^2 + \la_P P \Pt \right|^2
+ |c_P \avg{F_\phi^\dagger} + \la_P S|^2 (|P|^2 + |\Pt|^2)
\\
& \qquad
+ |\avg{F_\phi}|^2 \left( \sfrac 12 c_S S^2 + c_P P \Pt \right) + \hc
\eal\eeq
The potential is quadratic in $P$, $\Pt$, so we look for a minimum
with $\avg{P} = \avg{\Pt} = 0$.
In the appendix, we minimize the potential for real couplings and
VEVs.
We show that the global minimum preserves $CP$ for
\beq
c_S < 0
\eeq
and the we obtain
\beq[Smin]
\avg{S} &= - \frac{\avg{F_\phi}}{2\la_S}
\left( 3 c_S + \sqrt{c_S (c_S - 8)} \right),
\\
\Avg{\frac{F_S}{S}} &= \frac{\avg{F_\phi}}{4} \left(
-c_S + \sqrt{c_S (c_S - 8)} \right).
\eeq
This gives rise to a mass term for $P$, $\Pt$ that can be
conveniently written as
\beq[Pmasss]
\De \scr{L} = \myint d^2 \th\, \phi \scr{M} P \Pt + \hc,
\eeq
where
\beq
\scr{M} = M [1 + \th^2 r \avg{F_\phi}],
\eeq
In this parameterization $r \ne 0$
parameterizes the deviation from a supersymmetric threshold,
\ie $r = 0$ gives a pure anomaly-mediated spectrum below the messenger
scale.
The model of Nelson and Weiner has $r = -2$.
We then have
\beq
M &= c_P (1 + X)  \avg{F_\phi},
\\
r &= -\frac{2 + \frac 14 X \left( c_S + 4 - \sqrt{c_S(c_S - 8)} \right)}
{1 + X},
\eeq
where
\beq
X = \frac{\la_P \avg{S}}{c_S \avg{F_\phi}}
= -\frac{\la_P}{2 c_P \la_S} \left( 3 c_S + \sqrt{c_S(c_S - 8)} \right).
\eeq
This shows that all values of $M$ and $r$ are allowed,
since $1 + X$ can be small and have either sign.
(Note that this does not require any Yukawa couplings to be large.)
In order to avoid a negative mass eigenvalue for the scalars
$P$, $\Pt$ at the minimum, we require
\beq
|(r + 1) \avg{F_\phi}| < |M|.
\eeq

We now evaluate the threshold contributions to the standard
model fields due to the $P$ fields.
The general formulas can be obtained from the methods of
\Refs{GR,ALGR}.
The soft SUSY breaking terms can be parameterized by
higher superspace components of dimensionless couplings via
\beq
m_0^2 &= -\frac{\d}{\d \th^2} \frac{\d}{\d \bar{\th}^2} \ln Z,
\\
m_{1/2} &= \frac{1}{g} \frac{\d}{\d\th^2} g,
\\
\la A &= -2 \frac{\d}{\d \th^2} \la,
\eeq
where all couplings are taken to be real superfields.
In the present model, all SUSY breaking is contained in the conformal
compensator and the $P$, $\Pt$ mass term, so we have
\beq
\frac{\d}{\d\th^2} = \frac 12 \avg{F_\phi}
\left( r \frac{\d}{\d\ln M} - \frac{\d}{\d\ln\mu} \right).
\eeq
Note that this implies the presence of mixed anomaly-
and gauge-mediated terms for scalar masses, as first pointed out
in \Ref{PR}.
In this way, we can obtain expressions for the soft masses
at the scale $M$ in the effective theory where $P$ and $\Pt$
have been integrated out:
\beq[generalsoft]
m_0^2(M) &= \frac 14 \avg{F_\phi}^2 \biggl\{
- r^2 \frac{\d\ga'}{\d g'_i} \be'_i
+ 2 r (r + 1) \frac{\d\ga}{\d g_i} \be'_i
- (r + 1)^2 \frac{\d\ga}{\d g_i} \be_i
\biggr\},
\\
m_{1/2}(M) &= \frac{1}{g} \avg{F_\phi} \left[
r \be'_g - (r + 1)\be_g \right],
\\
\eql{generalA}
A(M)  &= -\frac{1}{\la} \avg{F_\phi} \left[
r \be'_\la - (r + 1) \be_\la \right].
\eeq
Here primed (unprimed) quantities refer to the theory above (below)
the scale $M$.
The anomalous dimensions are defined by
\beq
\be_i = \frac{\d g_i}{\d\ln\mu},
\qquad
\ga = \frac{\d\ln Z}{\d\ln\mu}.
\eeq
The expression for the scalar masses can be simplified in the
case of fields with no Yukawa couplings to messengers,
for which $\ga' = \ga$.
We then have
\beq[mscalarsimp]
m_0^2(M) &= m_{0\,{\rm AMSB}}^2
+ \frac 14 r (r + 2) \avg{F_\phi}^2
\frac{\d\ga}{\d g_i} \De\be_i,
\eeq
where
\beq
m_{0\,{\rm AMSB}}^2 = -\frac 14 \avg{F_\phi}^2 \frac{\d\ga}{\d g_i} \be_i.
\eeq
and $\De\be = \be' - \be$.
Similarly, we can write
\beq
m_{1/2}(M) &= m_{1/2\,\rm AMSB} + \frac{r}{g} \avg{F_\phi} \De\be_g
\\
\eql{Asimp}
A(M) &= A_{\rm AMSB} - \frac{r}{\la} \avg{F_\phi} \De\be_\la.
\eeq
These expressions explicitly display the fact
that the soft masses reduce to the
AMSB values in the limit $r \to 0$.
The scalar masses (but not gaugino masses and $A$ terms)
also reduce to their AMSB values for $r \to -2$,
as in the model of Nelson and Weiner.
In the generalized model, all
soft masses reduce to the
gauge-mediated values in the limit $r \to \infty$
with $r \avg{F_\phi}$ held fixed.
For general $r$, the SUSY breaking spectrum
in this model interpolates continuously between
anomaly mediation and gauge mediation with a messenger
scale of order $10\TeV$ (assuming all dimensionless couplings
are order unity).

As with the case of pure gauge- and anomaly-mediated SUSY breaking,
\Eqs{generalsoft}--\eq{generalA}
are leading order results in a power series with subleading
corrections suppressed by
$\scr{O}((\avg{F_\phi} / M^2)^2)$
and $\scr{O}((r \avg{F_\phi} / M^2)^2)$.
In the present class of models, it is natural to
have $M \sim \avg{F_\phi}, r\avg{F_\phi}$, where these effects
may be important.
They have been calculated for the case of pure gauge mediation,
where they are known to be numerically small unless
the SUSY breaking is tuned to be close to the instability
limit $F / M^2 \to 1$ \cite{GMSBcalc}.
Because these corrections are UV finite, they do not depend
on the regulator, and therefore depend on the conformal compensator
only through the superfield mass of the messengers (see \Eq{Pmasss}).
We can therefore use the results for gauge mediation with the replacement
$F/M^2 \to (r + 1) \avg{F_\phi}/M^2$.
Since the stability limit is $|(r + 1) \avg{F_\phi}/M^2| < 1$ here
as well, the corrections are small in the absence of fine tuning.

\section{The $\mu$ Problem}
In the context of AMSB, we cannot get a phenomenologically acceptable
Higgsino mass by adding a
$\mu$ term
\beq
\De \scr{L} = \myint d^2\th\, \mu \phi H_u H_d + \hc
\eeq
since this gives rise to $B \sim \avg{F_\phi} \sim 10\TeV$.
One possibility is the NMSSM, where the VEV of a singlet
gives the $\mu$ term.
However, it is nontrivial to get a negative mass-squared term
for the singlet.
Here we briefly discuss another possibility within the MSSM that
gives a more minimal model.

We consider a mechanism originally proposed by
Randall and Sundrum in \Ref{AMSB}.
We show that this mechanism can be made natural with
appropriate broken symmetries.
We add a term to the Lagrangian of the form
\beq[RSmu]
\De\scr{L}_{\rm RS}
= \myint d^2 \th\, c (Y + Y^\dagger) \frac{\phi^\dagger}{\phi}
H_u H_d + \hc
\eeq
Here we have included factors of $\phi$ by canonically normalizing
$H_{u,d}$ but not the field $Y$.
Expanding this out, we obtain the potential terms
\beq\bal
\De\scr{L}_{\rm RS}
&= \left[ -c |F_\phi|^2 (Y + Y^\dagger) + c (F_\phi^\dagger F_Y - \hc)
\right] H_u H_d + \hc
\\
&\qquad
+ \left[ c F_\phi^\dagger (Y + Y^\dagger) + c F_Y^\dagger
\right] \myint d^2 \th\, H_u H_d
+ \hc
\eal\eeq
We see that we can naturally get a vanishing $B \mu$ term at
tree level if
\beq
\avg{Y + Y^\dagger} = 0
\eeq
and all couplings and VEV's are real.
This is natural by $CP$ invariance,
and we then obtain an
effective $\mu$ term
\beq
\mu = c \avg{F_Y}.
\eeq
The $B\mu$ term is generated from AMSB, giving rise to a model with
only one additional parameter.

It is crucial that the $Y$ appears in the
combination $Y + Y^\dagger$.
This is natural if the field $Y$ is invariant under a shift
symmetry
\beq
Y \mapsto Y + i \la
\eeq
where $\la$ is a real constant.
We must also forbid a term of the form
\beq[muunwanted]
\De\scr{L} = \myint d^4\th\, c' \frac{\phi^\dagger}{\phi} H_u H_d
+ \hc
\eeq
The discrete $R$ symmetry
\beq
Y(\th) \mapsto -Y(i\th),
\quad
H_u(\th) \mapsto H_u(i\th),
\quad
H_d(\th) \mapsto -H_d(i\th)
\eeq
forbids the unwanted term \Eq{muunwanted}, and also has the
feature that the lowest component of $Y$ is odd, while
$F_Y$ is even.
\footnote{%
This symmetry also forbids unwanted couplings between the singlet
$S$ and the Higgs fields if $S$ transforms as
$S(\th) \mapsto -S(i\th)$.}
The VEV for $Y$ that we need
is therefore protected by this symmetry.
In order to make the Yukawa couplings invariant, the standard
model fields must also transform under the discrete symmetry,
\eg
\beq
u^{\rm c}(\th) \mapsto -u^{\rm c}(i\th),
\eeq
with all other fields even.

This shows that the term \Eq{RSmu} with $Y$ treated as a spurion
provides a viable $\mu$ term in AMSB that is natural
by symmetries.
Effectively, it justifies the inclusion of a running $\mu$ term
into the AMSB RG trajectory.
It does not explain why the $\mu$ term is the same size as other
SUSY breaking terms.
We leave this for future work.

\section{Spectrum and Phenomenology}
We now discuss the SUSY breaking spectrum that results from this model.
We assume that the messengers come in complete $SU(5)$
multiplets, so that the gauge coupling unification
in the MSSM is not an accident.
The simplest possibility is then that the messengers consist of $N$
copies of
${\bf 5} \oplus \bar{\bf 5}$.
For perturbative unification, we require $N \le 4$.
Under the standard model gauge group,
these decompose into a doublet and a triplet, each of which can
have different couplings $c_P$ and $\la_P$ (see \Eq{Ladd}).
These give rise to different values for $r$ for the doublet and
triplet messengers, and hence different SUSY breaking masses for
colored and uncolored superpartners.
We assume for simplicity that the $N$ messengers have the same
coupling (\eg\ there can be an unbroken $SU(N)$ symmetry in the messenger
sector).
This can be relaxed to obtain even more
general spectra.

For large $r$, the spectrum is close to that of gauge mediation.
However, because SUSY breaking is driven by anomaly mediation, the gravitino
mass is naturally of order $\avg{F_\phi}$, alleviating the gravitino problem.
This may not be large enough for large $r$, but
it is possible (and natural) to have masses for the gravitino
and other gravitational moduli that are
parametrically larger than $\avg{F_\phi}$ with SUSY breaking dominated
by anomaly mediation \cite{largegravitino}.

The simplest model is completely specified at high energies by
$M$, $F_\phi$, $r_2$, $r_3$, $N$, and $\mu$.
One parameter is eliminated by requiring that the Higgs VEV
takes its experimentally determined value, so this model has
four continuous and one discrete parameter.%
\footnote{The top quark Yukawa coupling is fixed by demanding that
the top quark mass has its measured value.}
Of these, the dependence on the messenger scale is only logarithmic,
since it just sets the scale for
the RG running down to the weak scale.
Explicit formulas for soft masses are presented in Appendix B.

\begin{figure}[htbp]
         \begin{center}
             \includegraphics[width=5.5in]{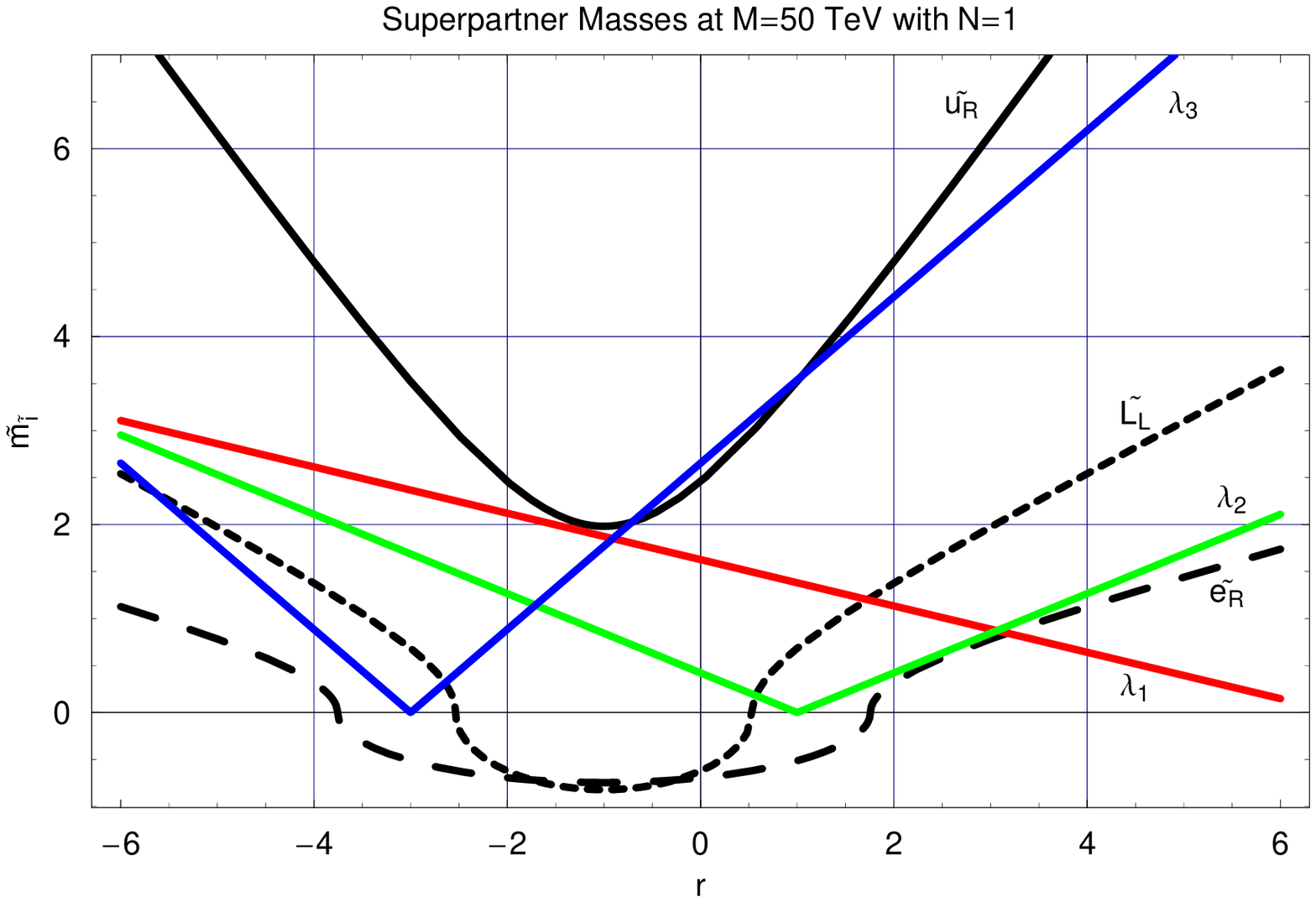}
             \includegraphics[width=5.5in]{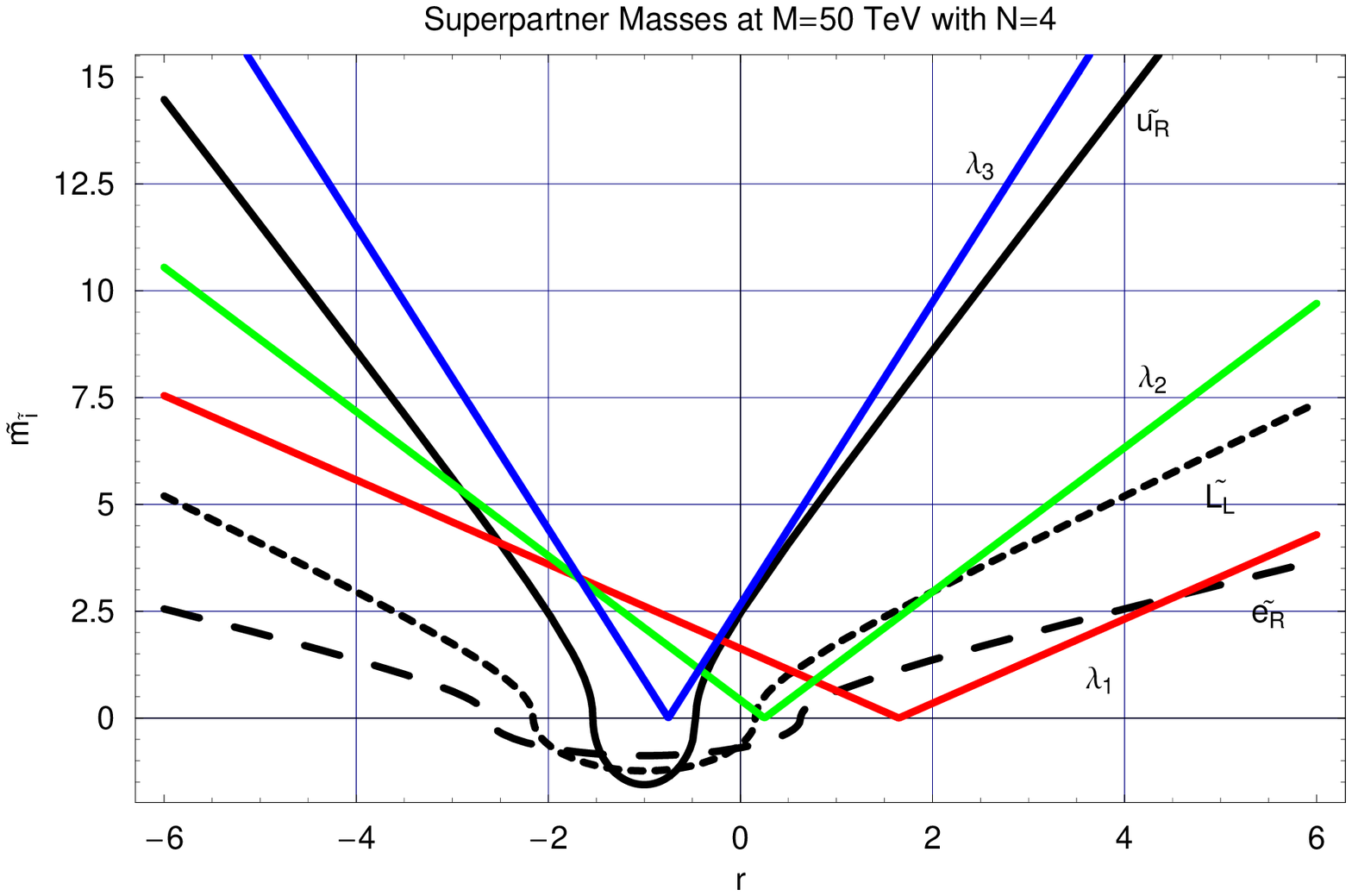}
             \caption{Spectrum of superpartner masses
             as a function of $r = r_2 = r_3$ for $M =50 \TeV$, and
             $N=1$ (top) and $N=4$ (bottom).
             For gaugino masses we plot $|M|$ and
             for scalar masses, we plot $ |m^2|^{1/2} \times \sgn(m^2)$.
             All masses are in units of $F_\phi /(16\pi^2)$.}
         \label{fig:samplespec}
         \end{center}
     \end{figure}

For illustration,
the spectrum of superpartner masses at the messenger scale
is shown in Fig.~\ref{fig:samplespec} as a function of $r = r_2 = r_3$ for
$M = 50$ TeV, for $N = 1$ and $N = 4$ respectively.
For $r < 0$ we can obtain positive slepton mass-squared
parameters, but the right-handed sleptons are lighter than
the bino, giving rise to charged slepton LSP.
We therefore focus our attention on $r > 0$.
The spectra are still
qualitatively similar to gauge- and
anomaly mediation in the sense that colored superpartners
are heavier than uncolored ones.
For example, obtaining positive slepton mass-squared parameters
requires $r \gsim 1$, which then implies
$m_{\tilde{q}} \gsim 5 m_{\tilde\ell}$.

\begin{figure}[btp]
         \begin{center}
             \includegraphics[width=5.5in]{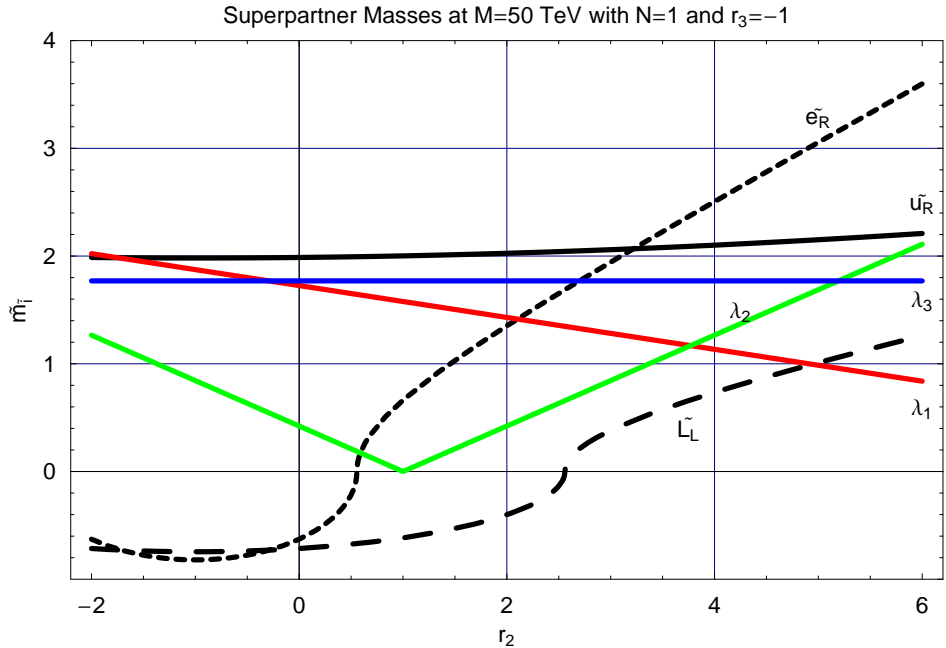}
             \caption{Spectrum of superpartner masses
             as a function of $r_2$ for $N = 1$, $M =50\TeV$, and $r_3 = -1$.
             For gaugino masses we plot $|M|$ and
             for scalar masses, we plot $ |m^2|^{1/2} \times \sgn(m^2)$.
             All masses are in units of $F_\phi /(16\pi^2)$.}
         \label{fig:r2r3spec}
         \end{center}
     \end{figure}

Quite different possibilities exist if $r_2 \ne r_3$.
In Fig.~\ref{fig:r2r3spec} we show an example spectrum with $N = 1$ and
$r_3 = -1$.
We again require $r > 0$ to avoid a slepton LSP.
We see that the spectrum is more degenerate, and the
$SU(2)_W$ contribution to superpartner masses is comparable
to $SU(3)_C$.
For $r_2 \gsim 2$, the superpartners charged under $SU(2)_W$
are the heaviest, followed by the gluino, then right-handed
scalars and the Bino.
Such spectra open up new regions of SUSY parameter space
that may be interesting to explore.
These spectra have a light stop, and therefore requires an
additional contribution to the Higgs quartic.
Possibilities include a ``fat'' Higgs \cite{fat}
or large $D$ terms from exotic gauge interactions \cite{Dterms}.

We give some representative points in parameter space in Table~\ref{spectrum},
assuming $r_2 = r_3$ for simplicity.
At the scale $M$ we evaluate the soft-breaking parameters
using \Eqs{generalsoft}--\eq{generalA}, and evolve them down using MSSM
RG equations to  the stop mass scale $m_{\tilde{t}}$.
(Since we have small mixing in the stop sector, we simply use the common
stop mass.)
At the scale $m_{\tilde{t}}$, we determine the $\mu$ parameter by
minimizing the one-loop effective potential.
This includes the largest
2-loop corrections to the effective potential
because we use a value of $y_t$ that includes 1-loop
QCD corrections \cite{EZ}.
We then add by hand the 2-loop QCD threshold corrections
to the higgs mass
$m^2_{h^0}$, although this is a small correction
($< 2\GeV$) for small stop mixing.

\begin{table}[h]
\begin{center}
\caption{Sample MSSM spectra. All masses are in GeV.
The main text gives the definition of fine-tuning.}
\label{spectrum}
\vspace{0.125in}
\begin{tabular}{|c|c|c|}
\hline & Point 1 & Point 2 \\
\hline
\hline $N$ &1 & 4 \\
\hline $r$ & 14.6&    6.45  \\
\hline $F_{\phi}$ & 7.19 TeV & 6.34 TeV \\
\hline $M$ & 201 TeV & 81 TeV \\
\hline $\mu$ & 485 &425 \\
\hline $\tan\beta$ & 17.1 & 17.7 \\
\hline
\hline $m_{h^0}$ & 115 & 115 \\
\hline $m_{\tilde{l}L}$ & 380  & 330 \\
\hline $m_{\tilde{l}R}$ & 190  & 150 \\
\hline $m_{\tilde{q}L}$ & 1220 & 1170 \\
\hline $m_{\tilde{u}R}$ & 1170 & 1130\\
\hline $m_{\tilde{d}R}$ & 1065 & 1120 \\
\hline $m_{\tilde{t}1}$ & 1070 & 1050 \\
\hline $m_{\tilde{t}2}$ & 1180 & 1150 \\
\hline $m_{\tilde{g}}$  & 880  & 1280 \\
\hline $m_{\tilde{\chi}_1^0}$ & 80 &   165 \\
\hline
\hline Tuning & 170 & 55 \\
\hline
\end{tabular}
\end{center}
\end{table}

The spectra given in Table~\ref{spectrum} satisfy all experimental
constraints.
The most severe constraint is the LEP Higgs mass bound
$m_{h^0}>114.4$ GeV.
Because we do not have large mixing in the
stop-sector, we require $m_{\tilde{t}}\sim 1$ TeV to satisfy
the Higgs mass bound, and the experimental
constraints on the sleptons and LSP are easily satisfied.
As we have large stop masses, these models are fine-tuned.

\begin{figure}[!h]
         \begin{center}
             \includegraphics[width=5in]{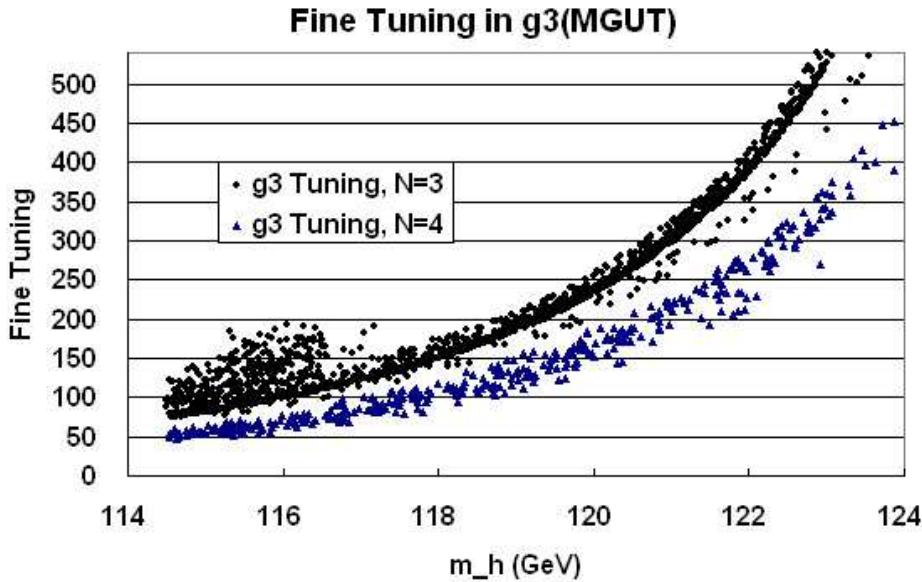}
             \caption{Fine-tuning in $g_3(M_{\mbox{\scriptsize GUT}})$ as a function of lightest Higgs mass $m_{h^0}$ for
             models with $r>0$ for $N=3$ and $4$.}
         \label{fig:g3tuning}
         \end{center}
     \end{figure}

We quantify the fine tuning by the sensitivity of the Higgs to
varying parameters at the GUT scale.
The Higgs mass is quadratically sensitive to the stop mass,
but this is not a fundamental parameter in this model.
The most sensitive fundamental parameter is $g_3(M_{\rm GUT})$,
so we define
\beq
\mbox{Fine tuning} \equiv \frac{g_3(M_{\rm GUT})}{v}
\frac{\partial v}{\partial g_3(M_{\rm GUT})}
= \frac{\partial\ln v}{\partial\ln  g_3(M_{\rm GUT}) }.
\eeq
Because the sensitivity is through the stop mass, the tuning
increases quadratically with the stop mass, while the lightest
Higgs mass increases only logarithmically.
This means that the fine tuning increases exponentially as a
function of the lightest Higgs mass.
This phenomenon is intrinsic to the MSSM, not just the present
model, and is illustrated in Fig.~\ref{fig:g3tuning}.
Note that the fine-tuning is somewhat less for a large number of messengers,
since QCD is non asymptotically free in this case,
and therefore the sensitivity to $g_3(M_{\rm GUT})$ is reduced.

\section{Conclusion}
We have constructed a well-motivated minimal model that naturally
breaks SUSY in a flavor-blind way with a messenger scale near 10~TeV.
The minimal model with one messenger has four continuous parameters and
one discrete parameter, and can give rise to spectra that are very
different from scenarios considered in the literature.
These include ``compact'' spectra with colored superpartners
close in mass to uncolored superpartners, a feature of the spectrum
that may help with SUSY naturalness.

\section*{Acknowledgements}
We thank Parul Rastogi and Haibo Yu for discussions.
This work is supported by the National Science Foundation
grant no. Phy-0354401 and the University of Maryland Center for
Particle and String Theory.

\appendix{Appendix A: Minimization of the Potential}
We minimize \Eq{theVS} assuming that all couplings are real.
It is useful to write the potential as
\beq
V = \la_S^2 \left\{
\left| \sfrac 12 S + \frac{c \avg{F_\phi}}{\la_S} \right|^2 |S|^2
+ \left( \frac{c_S \avg{F_\phi}}{\la_S} \right)^2
\left( \frac{1}{2c_S} S^2 + \hc \right)
\right\}
\eeq
and use units where $c_S \avg{F_\phi} / \la_S = 1$.
We see that the phase structure is completely determined by
the dimensionless parameter
\beq
\xi = \frac{1}{c_S}.
\eeq
Writing
\beq
\avg{S} = s e^{i\th},
\eeq
where $s$ and $\th$ are real, we have
\beq
\frac{V}{\la_S^2} = (1 + \xi \cos 2\th) s^2
+ s^3 \cos\th + \sfrac 14 s^4.
\eeq
This is stationary in $\th$ for
\beq[thetastationaryVS]
s = 0
\quad\hbox{\rm or}\quad
\sin\th = 0
\quad\hbox{\rm or}\quad
\cos\th = -\frac{s}{4\xi}.
\eeq
We consider these cases one at a time.

The case $\sin\th = 0$ is equivalent to $\avg{S} = s = \hbox{\rm real}$.
In that case, we find stationary points
\beq[sreal]
s = s_\pm = \sfrac 12 \left[ -3 \pm \sqrt{1 - 8 \xi} \right].
\eeq
Consistency therefore requires
$\xi < \sfrac{1}{8}$.
It is easy to check that
\beq
V(s_-) &< V(s_+), V(0) \quad\ \ \hbox{\rm for}\quad \xi < 0,
\\
V(0) &< V(s_-), V(s_+) \quad\, \hbox{\rm for}\quad 0 < \xi < \sfrac 18.
\eeq

It remains only to consider the third condition in
\Eq{thetastationaryVS}.
In this case, the stationary points are
\beq
s = \tilde{s}_\pm = \pm 2 \sqrt{\frac{\xi (\xi - 1)}{2\xi - 1}}.
\eeq
Reality of $s$ and $|\cos\th| \le 1$ are satisfied only if
\beq
\xi \ge 0.
\eeq
We have $V(\tilde{s}_+) = V(\tilde{s}_-)$, as we expect since
$CP$ is spontaneously broken.
We can check that
\beq
V(0) &< V(\tilde{s}_\pm) \quad\,\hbox{\rm for}\quad \xi < 1,
\\
V(\tilde{s}_\pm)  &< V(0) \quad\ \ \hbox{\rm for}\quad \xi > 1.
\eeq
We conclude that
\beq
\avg{S} = \begin{cases}
s_- & $\xi < 0$ \cr
0 & $0 < \xi < 1$ \cr
\tilde{s}_\pm e^{i\th_\pm} & $\xi > 1$,
\end{cases}
\eeq
where
\beq
\cos\th_\pm = \mp \sqrt{\frac{\xi - 1}{4\xi(2\xi - 1)}}.
\eeq
Restoring the units, we obtain the formulas used in the main text.

\appendix{Appendix B: Formulas for Soft Masses}
In this appendix, we give some explicit one-loop formulas for SUSY
breaking masses.
The beta functions for the MSSM gauge couplings with
$N_2$ doublets and $N_3$ triplets are
\beq
\be_i = \frac{b_i}{16\pi^2} g_i^3,
\eeq
where
\beq
b_3 &= -3 + N_3,
\\
b_2 &= 1 + N_2,
\\
b_1 &= 11 + N_2 + \sfrac 23 N_3.
\eeq
The one-loop anomalous dimensions are
\beq
\ga_{Q3} &= \frac{1}{16\pi^2} \left[
\sfrac{16}{3} g_3^2 + 3 g_2^2 + \sfrac 19 g_1^2 - 2 y_t^2 \right],
\\
\ga_{u3} &= \frac{1}{16\pi^2} \left[ \sfrac{16}{3} g_3^2 + \sfrac{16}{9}g_1^2 - 4 y_t^2 \right],
\\
\ga_{d3} &= \frac{1}{16\pi^2} \left[ \sfrac{16}{3} g_3^2 + \sfrac{4}{9}g_1^2 \right],
\\
\ga_{L} &= \frac{1}{16\pi^2} \left[ 3 g_2^2 + g_1^2 \right],
\\
\ga_{e} &= \frac{1}{16\pi^2} \left[ 4 g_1^2 \right],
\\
\ga_{Hu} &= \frac{1}{16\pi^2} \left[3 g_2^2+g_1^2-6y_t^2 \right],
\\
\ga_{Hd} &= \frac{1}{16\pi^2} \left[ 3 g_2^2+g_1^2 \right],
\eeq

For the quark fields of the first and second generation,
the top Yukawa coupling contribution should be dropped.
We do not include the other Yukawa couplings, since they are negligible
for small $\tan\be$.
The beta function for the top Yukawa coupling is
\beq
\be_{yt} = \frac{y_t}{16\pi^2} \left[ 6y_t^2-\sfrac{16}{3}g_3^2-3g_2^2-\sfrac{13}{9}g_1^2 \right].
\eeq
These formulas can be used to compute the MSSM soft masses using
\Eqs{mscalarsimp}--\eq{Asimp}
in the main text.
In the one-loop approximation, the contributions from the doublet
and triplet messengers just add, and we obtain \eg
\newcommand{\AMSBFac}{\frac{1}{2}\frac{\langle F_{\phi}\rangle^2}{(16\pi^2)^2}}
\beq
m^2_{\tilde{Q},{\rm AMSB}} &= \AMSBFac
\left[ 16 g_3^4 - 3 g_2^4 - \sfrac{11}{9} g_1^4 + 2y_t(16\pi^2\beta_{y_t})\right],
\\
\De m^2_{\tilde Q} &=
\AMSBFac \left[r_3 (r_3 + 2)N_3
\left( \sfrac{16}{3} g_3^4 + \sfrac{2}{27} g_1^4 \right)\right.
\nonumber\\
& \qquad\qquad
\left.+ r_2 (r_2 + 2)N_2
\left( 3 g_2^4 + \sfrac{1}{9} g_1^4 \right)\right],\\
m^2_{\tilde{u}_R,{\rm AMSB}} &=\AMSBFac\left[ 16g_3^4-\sfrac{176}{9}g_1^4+ 4y_t(16\pi^2\beta_{y_t})\right],\\
\Delta m^2_{\tilde{u}_R}
&=\AMSBFac\left[\sfrac{16}{3}g_3^4r_3(r_3+2)N_3
\right.\nonumber\\&\qquad\qquad\quad\left.
+\sfrac{16}{9}g_1^4\left(\sfrac{2}{3}r_3(r_3+2)N_3+r_2(r_2+2)N_2\right)\right],\\
m^2_{\tilde{d}_R,{\rm AMSB}} &=\AMSBFac\left[16g_3^4-\sfrac{44}{9}g_1^4\right],\\
\Delta m^2_{\tilde{d}_R} &=\AMSBFac\left[\sfrac{16}{3}g_3^4r_3(r_3+2)N_3
\right.\nonumber\\&\qquad\qquad\quad\left.
+\sfrac{4}{9}g_1^4\left(\sfrac{2}{3}r_3(r_3+2)N_3+r_2(r_2+2)N_2\right)\right],\\
m^2_{\tilde{L},{\rm AMSB}} &=\AMSBFac\left[-3g_2^4-11g_1^4\right],\\
\Delta m^2_{\tilde{L}} &=\AMSBFac\left[3g_2^4r_2(r_2+2)N_2
\right.\nonumber\\&\qquad\qquad\quad\left.
+g_1^4\left(\sfrac{2}{3}r_3(r_3+2)N_3+r_2(r_2+2)N_2\right)\right],\\
m^2_{\tilde{e}_R,{\rm AMSB}} &=\AMSBFac\left[-44 g_1^4\right],\\
\Delta m^2_{\tilde{e}_R} &=\AMSBFac\left[4g_1^4\left(\sfrac{2}{3}r_3(r_3+2)N_3+r_2(r_2+2)N_2\right)\right],\\
m^2_{H_u,{\rm AMSB}} &=\AMSBFac\left[-3g_2^4-11g_1^4+ 6y_t(16\pi^2\beta_{y_t})\right],\\
\Delta m^2_{H_u} &=\AMSBFac\left[3g_2^4r_2(r_2+2)N_2
\right.\nonumber\\&\qquad\qquad\quad\left.
+g_1^4\left(\sfrac{2}{3}r_3(r_3+2)N_3+r_2(r_2+2)N_2\right)\right],\\
m^2_{H_d,{\rm AMSB}} &=\AMSBFac\left[-3g_2^4-11g_1^4\right],\\
\Delta m^2_{H_d} &=\AMSBFac\left[3g_2^4r_2(r_2+2)N_2
\right.\nonumber\\&\qquad\qquad\quad\left.
+g_1^4\left(\sfrac{2}{3}r_3(r_3+2)N_3+r_2(r_2+2)N_2\right)\right].
\eeq
For the squarks of the first and second generation, we drop the top Yukawa coupling contribution.

The gaugino masses are given by
\beq
m_{\lambda_1}&=\frac{\langle F_\phi\rangle}{16\pi^2}\left(-11+\sfrac{2}{3}r_3N_3+r_2N_2\right),\\
m_{\lambda_2}&=\frac{\langle F_\phi\rangle}{16\pi^2}\left(-1+r_2N_2\right),\\
m_{\lambda_3}&=\frac{\langle F_\phi\rangle}{16\pi^2}\left(3+r_3N_3\right),
\eeq
where the first term in the parenthesis is the AMSB contribution while the remaining terms are contributions
from the doublet and triplet messengers.

\newpage

\end{document}